# Predicting Stress in Two-phase Random Materials and Super-Resolution Method for Stress Images by Embedding Physical Information

Tengfei Xing[1], Xiaodan Ren[1], Jie Li[1,2,*]

1. School of Civil Engineering, Tongji University, 1239 Siping Road, Shanghai 200092, China
2. The State Key Laboratory on Disaster Reduction in Civil Engineering, Tongji University, Shanghai 200092, China

*Corresponding author, E-mail: lijie@tongji.edu.cn

**Abstract**

Stress analysis is an important part of material design. For materials with complex microstructures, such as two-phase random materials (TRMs), material failure is often accompanied by stress concentration. Phase interfaces in two-phase materials are critical for stress concentration. Therefore, the prediction error of stress at phase boundaries is crucial. In practical engineering, the pixels of the obtained material microstructure images are limited, which limits the resolution of stress images generated by deep learning methods, making it difficult to observe stress concentration regions. Existing Image Super-Resolution (ISR) technologies are all based on data-driven supervised learning. However, stress images have natural physical constraints, which provide new ideas for new ISR technologies. In this study, we constructed a stress prediction framework for TRMs. First, the framework uses a proposed Multiple Compositions U-net (MC U-net) to predict stress in low-resolution material microstructures. By considering the phase interface information of the microstructure, the MC U-net effectively reduces the problem of excessive prediction errors at phase boundaries. Secondly, a Mixed Physics-Informed Neural Network (MPINN) based method for stress ISR (SRPINN) was proposed. By introducing the constraints of physical information, the new method does not require paired stress images for training and can increase the resolution of stress images to any multiple. This enables a multiscale analysis of the stress concentration regions at phase boundaries. Finally, we performed stress analysis on TRMs with different phase volume fractions and loading states through transfer learning. The results show the proposed stress prediction framework has satisfactory accuracy and generalization ability.

## 1. Introduction

The microstructure of two-phase random materials (TRMs) can be characterized by two distinct phases distributed in a random or stochastic manner[1], exhibiting complex behaviors that are challenging to capture with traditional analytical methods. TRMs are prevalent in the microstructures of biological and biomimetic materials, such as synthetic bone materials designed to mimic human cancellous bone[2], the air nanostructures found in male eastern bluebird feathers[3], and battery electrodes[4]. Accurately calculating the mechanical and physical properties of such materials is crucial for designing advanced composite materials, optimizing structural performance, and predicting failure mechanisms. In the processing-structure-property (PSP) paradigm of material design, whether it involves forward design optimization of materials[5] or the establishment of inverse mapping relationships using deep learning methods[6], rapid analysis of material stress is an essential and crucial step.

In stress analysis, traditional numerical methods, such as the finite element method (FEM), often struggle with accurately modeling material distributions and complex geometric regions. Although techniques like adaptive meshing and local refinement can enhance FEM performance, they introduce significant computational overhead. Therefore, many researchers use deep learning techniques for the rapid global stress prediction of material microstructures. Bhaduri et al. [7] used a U-net architecture to predict the stress in fiber-reinforced composites and successfully demonstrated the method's generalization capability across multiscale microstructures through transfer learning. Yuan et al. [8] further focused on the plastic behavior of fiber-reinforced composites and designed a transformer-based network architecture for stress prediction during the plastic phase. Buehler et al. [9] achieved bidirectional prediction of material microstructures and stress fields using generative adversarial networks (GAN). Jadhav et al. [10] proposed the StressD model which predicts normalized stress images using a U-net-based denoising diffusion model and then scales the normalized stress using an auxiliary model. Maurizi et al. [11] used graph neural networks (GNN) to examine three mechanical states of fiber composites: plasticity, wrinkling of layer interfaces, and buckling instability. The results demonstrate the effectiveness of GNN in predicting stress under different mechanical states of materials. These studies indicate that most deep learning methods can accurately and quickly compute the stress state of material microstructures. However, most existing studies exhibit significantly higher stress prediction errors at phase boundaries compared to other regions. The interface effect indicates that the interface is the connecting region between different phases, and its strength directly affects the overall mechanical properties of the material. Poor interfacial bonding can lead to stress concentration, promoting crack propagation along the interface and resulting in a decline in material performance [12-14]. Therefore, there is a need to develop a rapid prediction method that can accurately reduce stress errors at phase boundaries.

Due to the small scale of TRM microstructures, it is difficult to obtain high-resolution images of the microstructure in practical problems. Additionally, the resolution of the finally predicted stress image depends entirely on the dataset, which limits the applicability of the above deep learning-based stress prediction methods. Considering that the failure of composite materials is often accompanied by stress concentration when we need to focus on the stress concentration areas of the material microstructure, it is necessary to obtain dense stress data in that area, which requires an increase in the resolution of the stress image. Existing deep learning-based Image Super-Resolution (ISR) methods, such as Single Image

Super-Resolution (SISR)[15], Generative Adversarial Networks for Super-Resolution (SRGAN)[16], etc., are closely related to the low-resolution images and the high-resolution images in the dataset. However, if a large number of high-resolution stress images are generated using FEM, it will sacrifice a lot of time, which is contrary to the original intention of the efficiency of deep learning. Stress images naturally satisfy the constraints of physical information[17], which is different from general images. This provides the possibility for the stress image super-resolution technique based on the Mixed Physics-Informed Neural Network (MPINN).

Existing studies have demonstrated the effectiveness of MPINN in computing the stress of heterogeneous materials. For example, Rezaei et al. [18] used MPINN to solve the elastic and steady-state diffusion problems of heterogeneous materials. Ren et al. [19] used MPINN to address the elastic and plasticity problems of two-phase random materials and demonstrated the generalization ability of MPINN under different material loading states through transfer learning. However, the accuracy of stress prediction comes at the cost of a large number of training points. Therefore, there is a need to develop an MPINN method that can significantly reduce the number of selected points.

In this study, we established a stress analysis framework for TRMs. First, to avoid expensive experimental costs, we generated random fields using a stochastic harmonic function (SHF) and horizontally sliced the random fields to obtain TRM microstructure images. Then, we proposed a Multiple Compositions U-net (MC U-net) architecture, which fully considers the phase contour information of TRM microstructures and can be used to preliminarily predict the stress of low-resolution TRM microstructures. Additionally, we compared the effects of different concatenation positions and widths of phase interface information on the prediction ability of MC U-net. Furthermore, we proposed a stress ISR method based on MPINN (SRMPINN), in which the training process only requires the stress image from MC U-net and incorporating physical information. Meanwhile, we discussed the impact of loss function weights and image resolution magnification on the performance of SRMPINN in terms of error. Finally, to demonstrate the generalization ability of the proposed stress analysis framework, we performed transfer learning on TRMs with different phase volume fractions and loading states. The overall process of the above framework is illustrated in **Fig. 1**.

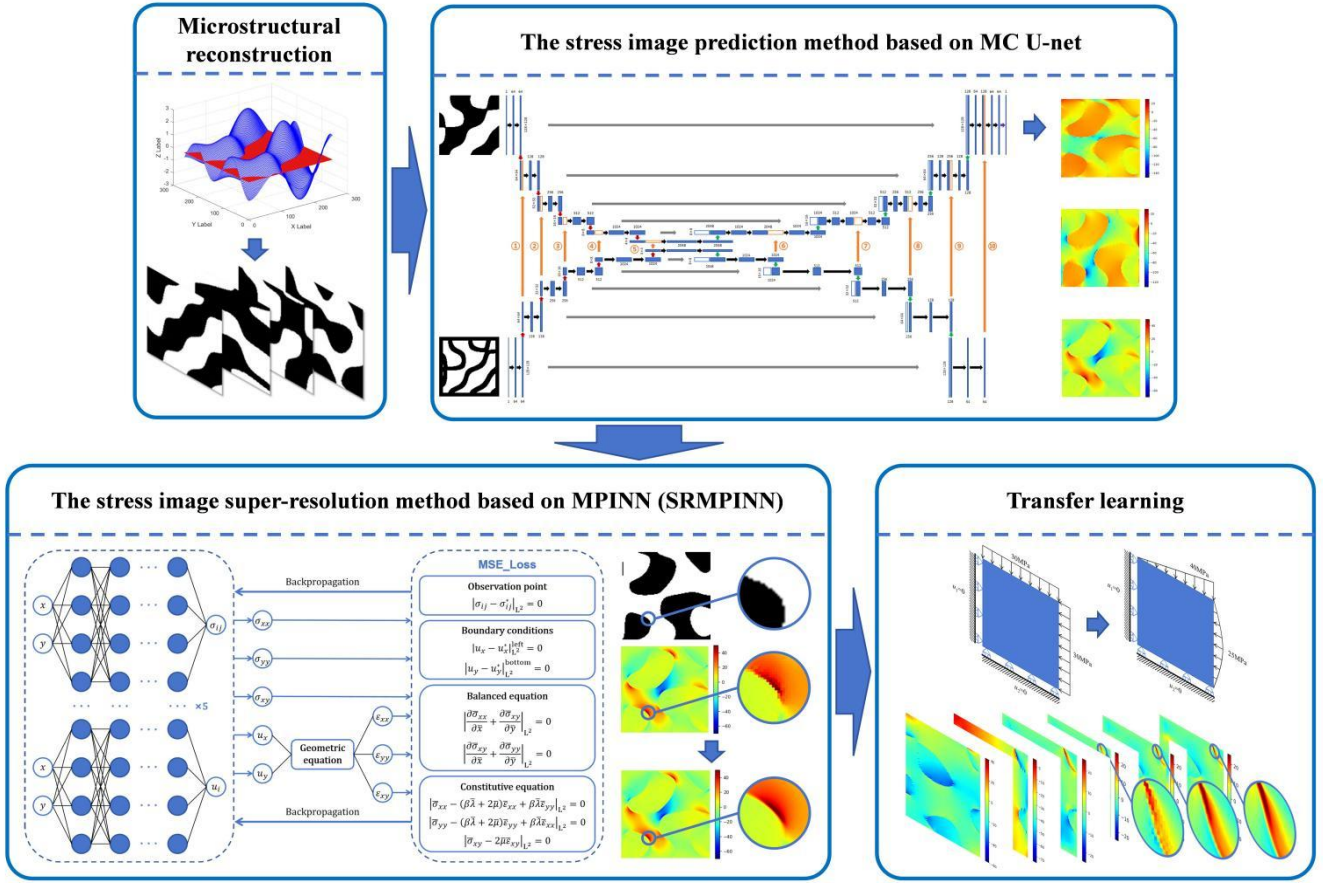

**Fig. 1.** Stress analysis framework for two-phase random materials.

This manuscript is organized as follows: Section 2 provides a detailed presentation of the stress analysis framework for TRMs, including the reconstruction method of the TRM microstructures based on the random field segmentation using stochastic harmonic functions, the stress image prediction method based on MC U-net, and the stress image super-resolution method based on MPINN (SRMPINN). Section 3 discusses the results of the proposed framework and compares them with some existing methods. Section 4 extends the proposed framework, by using transfer learning to analyze the stress of TRM microstructures under different phase volume fractions or different loading states. Section 5 provides conclusions.

## 2. Methodology
### 2.1. Microstructural reconstruction of two-phase random materials

The microstructure of random heterogeneous materials can be viewed as a probability space, where each microstructure sample can be generated by a specific random field.[20,21] Therefore, it is possible to generate microstructures by segmenting high-dimensional Gaussian random fields with the same power spectral density function. In this study, high-dimensional Gaussian fields are generated by stochastic harmonic function (SHF). The method can approximate the two-point correlation function of the original microstructure with fewer summation terms compared to conventional approaches. The stochastic harmonic function can be expressed as[22]

$$Y^{SHF}(x_1, x_2) = \sum_{i=1}^{N_1} \sum_{j=1}^{N_2} A(\omega_{1,i}, \omega_{2,j}) \left[ \cos(\omega_{1,i} x_1 + \omega_{2,j} x_2 + \varphi_{ij}^{I}) + \cos(-\omega_{1,i} x_1 + \omega_{2,j} x_2 + \varphi_{ij}^{II}) \right] \quad (1)$$

where $\varphi_{ij}^{I}$ and $\varphi_{ij}^{II}$ $(i=1,2,...,N_1; j=1,2,...,N_2)$ denote the random phase independently identically distributed over $(0, 2\pi]$. Assume the concerned wavenumber domain is $\Omega^{SHF} = [\omega_1^L, \omega_1^U) \times [\omega_2^L, \omega_2^U)$, which was divided into non-overlapping sub-domains $\Omega_{i,j}^{SHF} = [\omega_{1,i}^L, \omega_{1,i}^U) \times [\omega_{2,j}^L, \omega_{2,j}^U)$. The superscripts L and U represent the lower and upper bounds of the truncated wavenumber domain in the corresponding directions. $\omega_{1,i}$ and $\omega_{2,j}$ $(i=1,2,...,N_1; j=1,2,...,N_2)$ are uniformly distributed in the intervals $[\omega_{1,i}^L, \omega_{1,i}^U)$ and $[\omega_{2,j}^L, \omega_{2,j}^U)$ respectively. The amplitude of each item can be expressed as

$$A(\omega_{1,i}, \omega_{2,j}) = \sqrt{4S(\omega_{1,i}, \omega_{2,j})(\omega_{1,i}^U - \omega_{1,i}^L)(\omega_{2,j}^U - \omega_{2,j}^L)} \quad (2)$$

where $S(\omega_{1,i}, \omega_{2,j})$ is the spectral density function of the Gaussian random field.

Assuming the autocorrelation function of the Gaussian stationary random field is exponential, its parameters can be determined by the two-point correlation function.[23] The autocorrelation function can be expressed as

$$R(\delta_1, \delta_2) = \exp\left[ -\left(\frac{\delta_1}{c_1}\right)^2 - \left(\frac{\delta_2}{c_2}\right)^2 \right] \quad (3)$$

where $c_i$ $(i=1,2)$ is the scaling factor associated with the scales of fluctuation or correlation lengths in the different directions.

According to the Wiener-Khinchin theorem, the power spectral density function can be expressed as

$$S(\omega_1, \omega_2) = \frac{c_1 c_2}{4\pi} \exp\left[ -\left(\frac{c_1 \omega_1}{2}\right)^2 - \left(\frac{c_2 \omega_2}{2}\right)^2 \right] \quad (4)$$

After obtaining the high-dimensional Gaussian field samples, the desired microstructure image with a size of 128×128 is obtained by horizontal slicing, as shown in **Fig. 2**. The generated microstructure is statistically equivalent to the real material microstructure, and it represents the actual material for subsequent stress prediction and super-resolution.

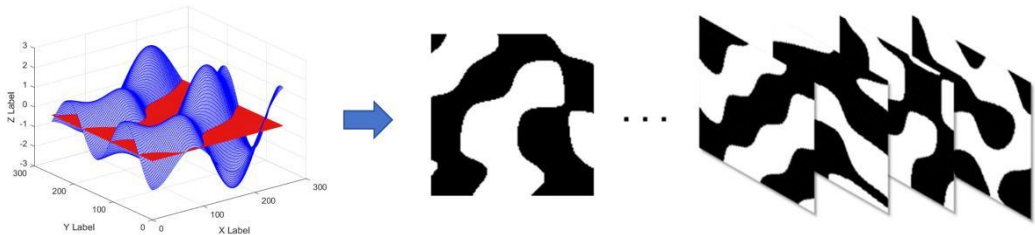

**Fig. 2.** Microstructure of two-phase random materials.

## 2.2. Stress prediction

The U-net architecture is effective in extracting the latent abstract information of images and decoding it. [24] Over the years, various improved forms based on U-net have been derived. Among them, Attention U-net is one of the more widely used improvements.[25] In the work, we have made certain modifications to the Attention U-net to better predict the stress of the microstructure of two-phase random materials, as shown in **Fig. 3**.

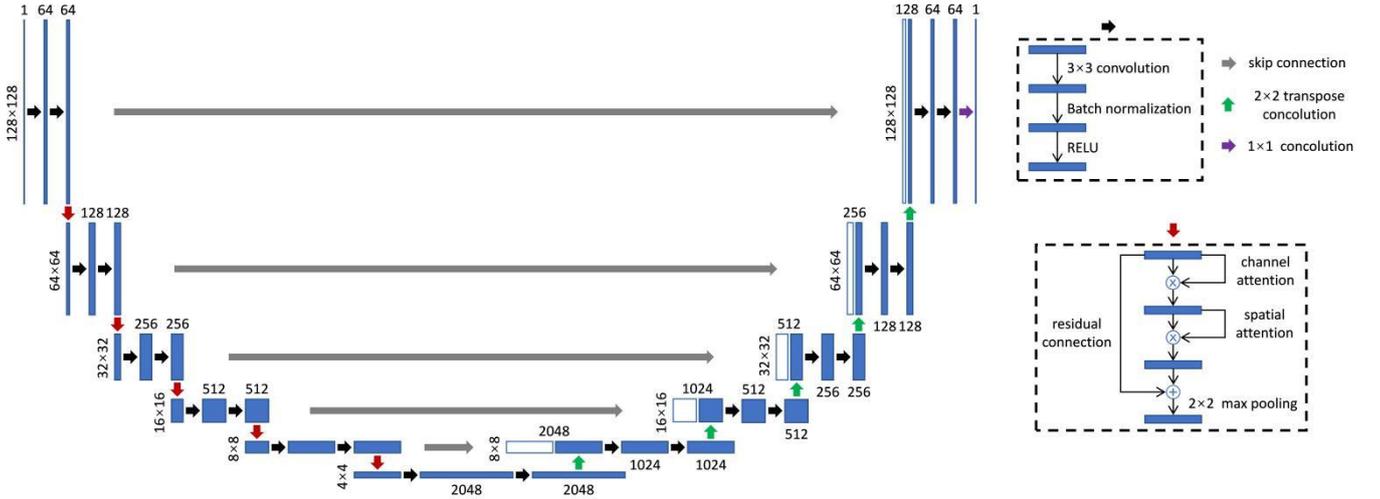

**Fig. 3.** Attention U-net architecture.

The encoder consists of 6 identical modules, each including (1) Two 3×3 convolutions, batch normalization, and rectified linear unit (ReLU) operation. (2) Downsampling, including channel attention, spatial attention, residual connection, and 2×2 max pooling. The 6th encoder module does not include the downsampling operation. The decoder consists of 5 identical modules, each including (1) Upsampling, essentially a 2×2 transpose convolution. (2) Skip connection, to utilize the information lost during downsampling in the decoding process as much as possible. (3) Two 3×3 convolutions, batch normalization, and rectified linear unit (ReLU) operation. The 5th decoder module additionally includes a 1×1 convolution layer at the end, which maps the 64-channel decoder output to a single channel. Additionally, it is worth noting that our Attention U-net has only one feature layer in both the input and output images.

The process of stress calculation using the Attention U-net can be expressed as

$$\boldsymbol{\sigma}_{NN} = A(\boldsymbol{x}_b) \tag{5}$$

where $\boldsymbol{\sigma}_{NN}$ is the stress value predicted by the neural network, $\boldsymbol{x}_b$ is the global information of the microstructure, and $A(\cdot)$ represents the operations of the Attention U-net.

Accurately predicting the stress values in the interface regions of different phases of two-phase random materials using only Attention U-net is challenging, because the stress variations in these regions exhibit strong nonlinearity. In order to solve the above problem, we drew some inspiration from the modeling of wind loads. In the modeling of wind loads, wind speed is divided into mean wind speed and fluctuating wind speed, which can be expressed as

$$v = v_a + v_d \tag{6}$$

where $v$ is the wind speed, $v_a$ is the mean wind speed, representing the average value of wind speed measured multiple times over a certain period. $v_d$ is the fluctuating wind speed, representing the random variations in wind speed and significant short-term fluctuations.

Similarly, we consider the final predicted stress images as consisting of two parts: global stress and phase interface stress. The phase interface stress reflects significant variations in the stress within the phase interface region to a certain extent. Global stress and phase interface stress are calculated based on the global information and phase interface information of the microstructure, respectively. This can be expressed as

$$\sigma_{NN} = f\left(G(x_b) \odot I(x_i)\right) \tag{7}$$

where $G(\cdot)$ is the operation on $x_b$ before tensor concatenation, $x_i$ and $I(\cdot)$ are the phase interface information of the microstructure and the operation on $x_i$ before tensor concatenation, respectively. $\odot$ is the tensor concatenation operator. $f(\cdot)$ is the operation after the concatenation of $G(x_b)$ and $I(x_i)$. Therefore, we extracted the phase interface information of the TRM microstructure, as shown in **Fig. 4**.

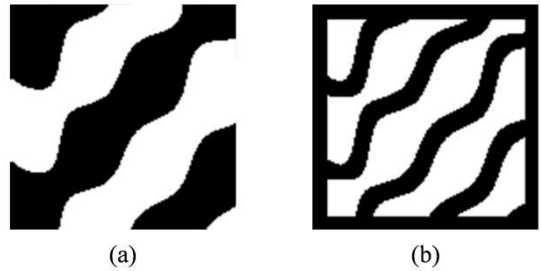

**Fig. 4.** (a) Microstructure of two-phase random materials. (b) Corresponding phase interface information.

Based on **Eq. (7)**, we proposed the Multiple Compositions U-net (MC U-net) architecture to address a class of problems characterized by strong nonlinearity due to boundaries, as illustrated in **Fig. 5**. The indices in the figures (①, ②, …, ⑩) indicate the selectable tensor concatenation positions. In other words, we incorporated the phase interface information of the latent space into the latent space of TRMs, enabling the stress predictions of the MC U-net to account for more comprehensive information. The Mean Squared Error is used as the loss function for weight training, which can be expressed as

$$\text{Mean Squared Error} = \frac{1}{N}\sum_{i=1}^{N}\sum_{j=1}^{M}\left(y_{ij}-\hat{y}_{ij}\right)^2 \tag{8}$$

where $N$ is the number of stress image samples, $M$ is the number of elements in each sample, $y_{ij}$ and $\hat{y}_{ij}$ denote the true value and predicted value, respectively, of the $j$-th element in the $i$-th sample.

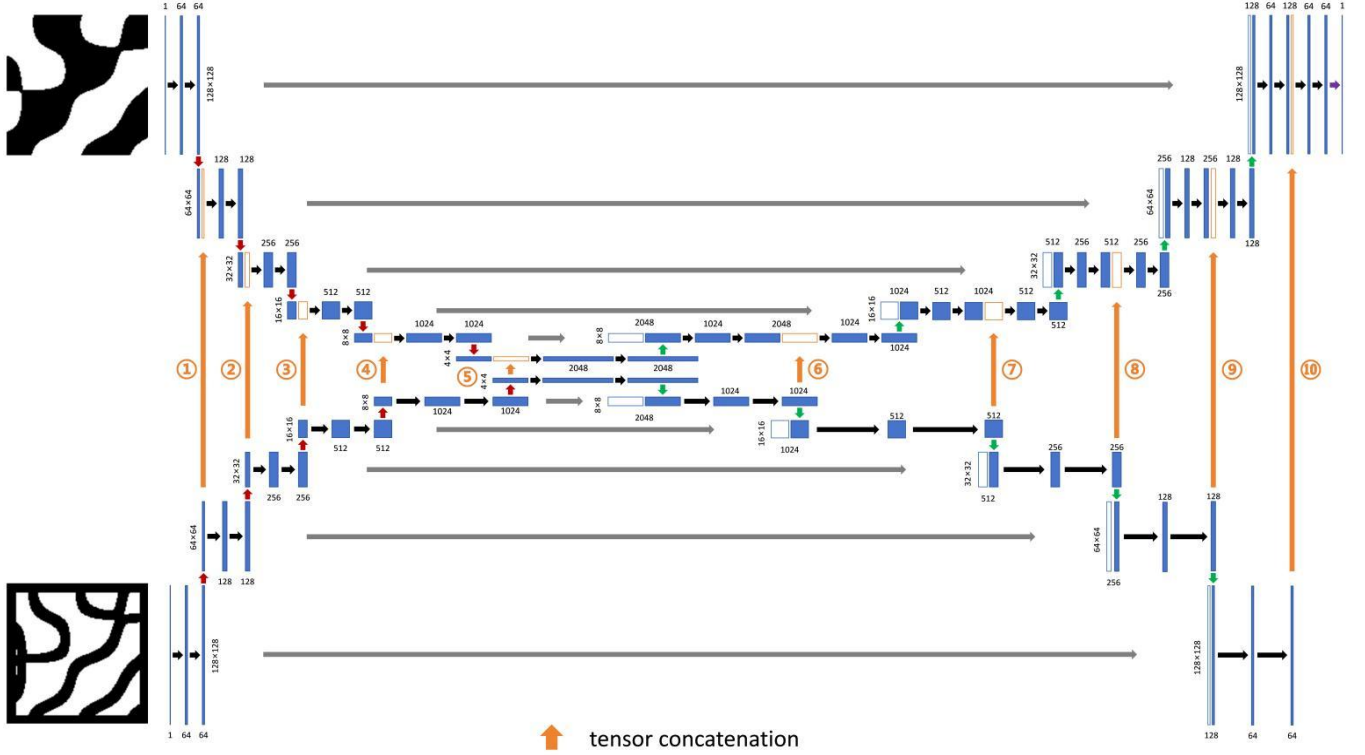

**Fig. 5.** MC U-net architecture.

**2.3. Stress image super-resolution**

In the field of image processing, a feedforward neural network (FFN) is not as effective as a convolutional neural network (CNN). However, most ISR methods based on CNN architectures require both low-resolution and high-resolution images to be provided, and the magnification factor of the resolution is closely related to the dataset. In contrast, for images with strong physical information constraints, such as the stress images in this study, the stress value at each pixel theoretically adheres strictly to the constraints of the corresponding partial differential equations. Because of these physical information constraints, using the FFN-based MPINN for ISR has proven to be highly successful, as discussed in Section 3. Notably, stress image super-resolution based on MPINN is not a fully data-supervised learning method, as it does not require any data annotations. During the weight training process, only low-resolution stress images are needed. In the prediction phase, images with any magnification factor can be obtained by simply increasing the dimension of the FFN input vector. This differs from ISR based on CNN architectures.

In this study, an MPINN architecture was designed using elastic materials as an example, as shown in **Fig. 6**. The architecture comprises five parallel FFNs serving as the main network structure, with each FFN outputting normal stresses, shear stresses, and displacements. Due to the geometric equation in **Eq. (9)**, the output displacement and output strain are equivalent. However, the choice of outputting displacement here is because displacement boundary conditions are easier to constrain and the architecture can reduce the number of parameters in the neural network compared to the three output values corresponding to strain. The loss function consists of four parts: observation point constraints (the stress values obtained through MC U-net), displacement boundary conditions, the equilibrium equation **Eq. (10)**, and the constitutive relation **Eq. (11)**.

$$\varepsilon = \frac{1}{2}\left(\nabla \otimes \boldsymbol{u} + \boldsymbol{u} \otimes \nabla\right) \tag{9}$$

$$\nabla \cdot \boldsymbol{\sigma} = 0 \tag{10}$$

$$\boldsymbol{\sigma} - \lambda \mathrm{tr}(\boldsymbol{\varepsilon})\mathbf{I} - 2\mu\boldsymbol{\varepsilon} = 0 \tag{11}$$

where $\nabla$ is the Hamiltonian operator, $\boldsymbol{u}$ is the displacement vector, $\boldsymbol{\varepsilon}$ and $\boldsymbol{\sigma}$ are the Cauchy strain and stress, respectively, $\mathrm{tr}(\cdot)$ is the trace of a matrix, $\mathbf{I}$ is the second-order identity tensor. $\lambda$ and $\mu$ are Lamé coefficient, related to the material's Young's modulus $E$ and Poisson's ratio $v$:

$$\lambda = \frac{Ev}{(1+v)(1-2v)}, \quad \mu = \frac{E}{2(1+v)} \tag{12}$$

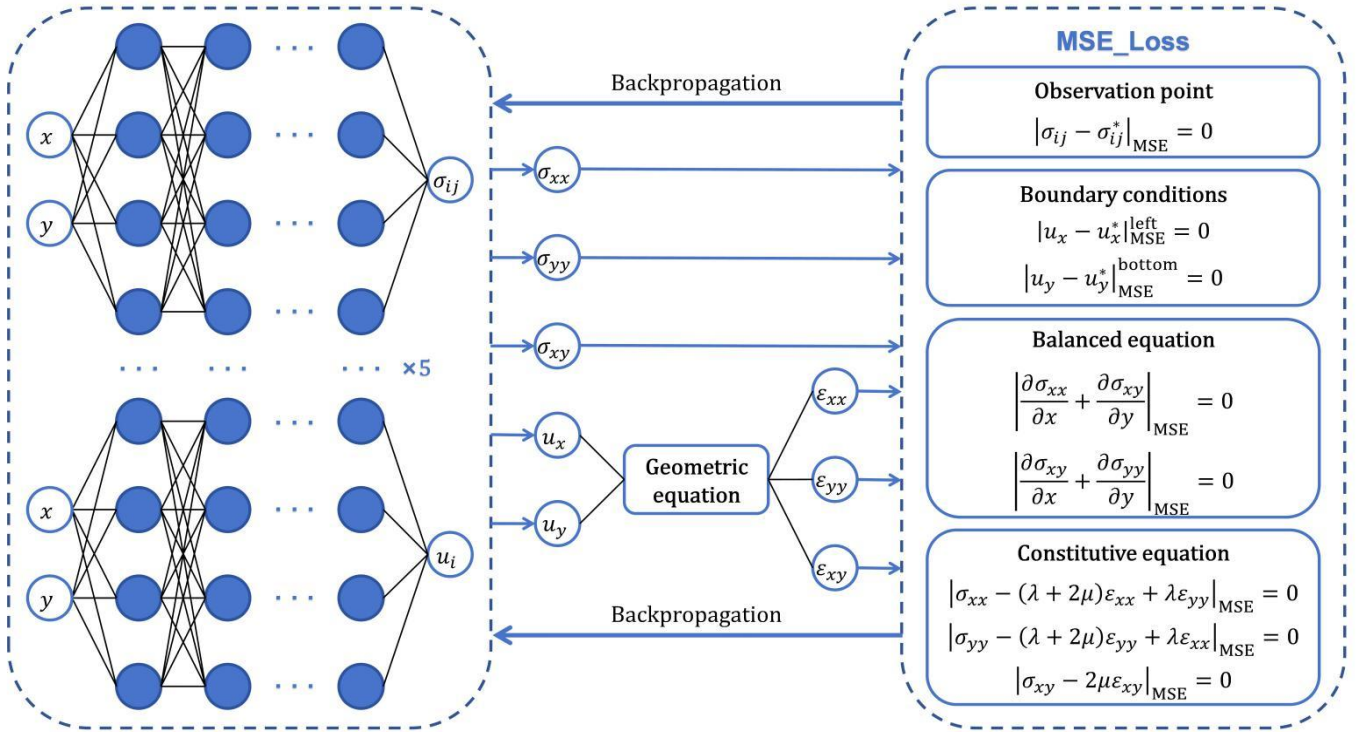

**Fig. 6.** MPINN architecture.

To avoid multi-scale loss caused by the significant difference in the order of magnitude of different parameters, it is necessary to dimensionless the parameters related to spatial coordinates in the constraint of physical information,[26]

$$\bar{x} = \frac{x}{L}, \quad \bar{y} = \frac{y}{L}, \quad \bar{\lambda} = \frac{\lambda}{\lambda_c}, \quad \bar{\mu} = \frac{\mu}{\mu_c}, \quad \bar{u} = \frac{u}{u_c}, \quad \bar{\sigma} = \frac{\sigma}{\sigma_c} \tag{13}$$

where $L$ is the maximum value of the spatial coordinate interval, $\lambda_c$ and $\mu_c$ are the maximum values of $\lambda$ and $\mu$ within the spatial range, respectively:

$$\lambda_c = \max_{x,y} \lambda, \quad \mu_c = \max_{x,y} \mu \tag{14}$$

To calculate $u_c$ and $\sigma_c$, we first substitute **Eq. (9)** into **Eq. (11)**, yielding:

$$\boldsymbol{\sigma} = \lambda \nabla \cdot \boldsymbol{u} \mathbf{I} + \mu (\nabla \otimes \boldsymbol{u} + \boldsymbol{u} \otimes \nabla) \tag{15}$$

Replace the parameters related to spatial coordinates with dimensionless parameters:

$$\sigma_c \bar{\boldsymbol{\sigma}} = \lambda_c \bar{\lambda} u_c L^{-1} \bar{\nabla} \cdot \bar{\boldsymbol{u}} \mathbf{I} + \mu_c \bar{\mu} u_c L^{-1} \left( \bar{\nabla} \otimes \bar{\boldsymbol{u}} + \bar{\boldsymbol{u}} \otimes \bar{\nabla} \right) \tag{16}$$

Extract the common factor $\mu_c u_c L^{-1}$ on the right side of **Eq. (16)**, we obtain,

$$\sigma_c \bar{\boldsymbol{\sigma}} = \mu_c u_c L^{-1} \left[ \bar{\nabla} \cdot \bar{\boldsymbol{u}} \beta \bar{\lambda} \mathbf{I} + \bar{\mu} \left( \bar{\nabla} \otimes \bar{\boldsymbol{u}} + \bar{\boldsymbol{u}} \otimes \bar{\nabla} \right) \right] \tag{17}$$

where $\beta$ equals $\lambda_c / \mu_c$. The scaling factor of the Cauchy stress tensor $\sigma_c$ equals $\mu_c u_c L^{-1}$.

The Navier-Cauchy equation for static problems can be expressed as,

$$\nabla \otimes \left[ (\lambda + \mu) \nabla \cdot \boldsymbol{u} \right] + \nabla \cdot (\mu \nabla \otimes \boldsymbol{u}) = 0 \tag{18}$$

Replace the parameters related to spatial coordinates with dimensionless parameters:

$$u_c L^{-2} \bar{\nabla} \otimes \left[ \left( \lambda_c \bar{\lambda} + \mu_c \bar{\mu} \right) \bar{\nabla} \cdot \bar{\boldsymbol{u}} \right] + u_c L^{-2} \bar{\nabla} \cdot \left( \mu_c \bar{\mu} \bar{\nabla} \otimes \bar{\boldsymbol{u}} \right) = 0 \tag{19}$$

Extract the common factor $\mu_c u_c L^{-2}$ on the left side of **Eq. (19)**, we obtain,

$$\mu_c u_c L^{-2} \left\{ \bar{\nabla} \otimes \left[ \left( \beta \bar{\lambda} + \bar{\mu} \right) \bar{\nabla} \cdot \bar{\boldsymbol{u}} \right] + \bar{\nabla} \cdot \left( \bar{\mu} \bar{\nabla} \otimes \bar{\boldsymbol{u}} \right) \right\} = 0 \tag{20}$$

Setting the coefficient of this equation to $r$ (constant), we obtain the scaling factor for the displacement vector $u_c = r \mu_c^{-1} L^2$.

The dimensionless MPINN loss function can be expressed as,

$$\mathscr{L}_{PINN} = \lambda_{OP} \mathscr{L}_{observation} + \lambda_{PI} \mathscr{L}_{physical\_information} = \lambda_{OP} \mathscr{L}_{observation} + \lambda_{PI} \left( \mathscr{L}_{boundary} + \mathscr{L}_{balanced} + \mathscr{L}_{constitutive} \right) \tag{21}$$

with

$$\mathscr{L}_{observation} = \left| \sigma_{ij} - \sigma_{ij}^* \right|_{MSE} \tag{22}$$

$$\mathscr{L}_{boundary} = \left| u_x - u_x^* \right|_{MSE}^{left} + \left| u_y - u_y^* \right|_{MSE}^{bottom} \tag{23}$$

$$\mathscr{L}_{balanced} = \left|\frac{\partial \bar{\sigma}_{xx}}{\partial \bar{x}} + \frac{\partial \bar{\sigma}_{xy}}{\partial \bar{y}}\right|_{MSE} + \left|\frac{\partial \bar{\sigma}_{xy}}{\partial \bar{x}} + \frac{\partial \bar{\sigma}_{yy}}{\partial \bar{y}}\right|_{MSE} \quad (24)$$

$$\mathscr{L}_{constitutive} = \left|\bar{\sigma}_{xy} - 2\bar{\mu}\bar{\varepsilon}_{xy}\right|_{MSE} + \left|\bar{\sigma}_{xx} - (\beta\bar{\lambda} + 2\bar{\mu})\bar{\varepsilon}_{xx} - \beta\bar{\lambda}\bar{\varepsilon}_{yy}\right|_{MSE}$$
$$+ \left|\bar{\sigma}_{yy} - (\beta\bar{\lambda} + 2\bar{\mu})\bar{\varepsilon}_{yy} - \beta\bar{\lambda}\bar{\varepsilon}_{xx}\right|_{MSE} \quad (25)$$

where $|\cdot|_{MSE}$ is the mean squared error loss function, $\lambda_{OP}$ and $\lambda_{PI}$ are the weight of the observation points part and the physical information part of the loss function, respectively, $\bar{\varepsilon} = (\bar{\nabla}\otimes\bar{u} + \bar{u}\otimes\bar{\nabla})/2$ is the dimensionless Cauchy strain tensor.

## 3. Results

### 3.1. Problem setup and boundary conditions

In this section, the mechanical problems studied are focused on square TRMs with a length of 2mm. In the grayscale image of TRM microstructures, the elastic moduli of the black phase and the white phase are $2.06\times10^4$ MPa and $2.06\times10^5$ MPa, respectively; the Poisson's ratios of the black phase and the white phase are 0.2 and 0.4, respectively. The problem is two-dimensional and involves a biaxial compression scenario with Neumann boundary conditions for static loading. Specifically, the right side and the top side of the domain both experience the uniform pressure of 30MPa. **Fig. 7a** illustrates the problem setup and boundary conditions. The blue regions represent the TRM microstructures。

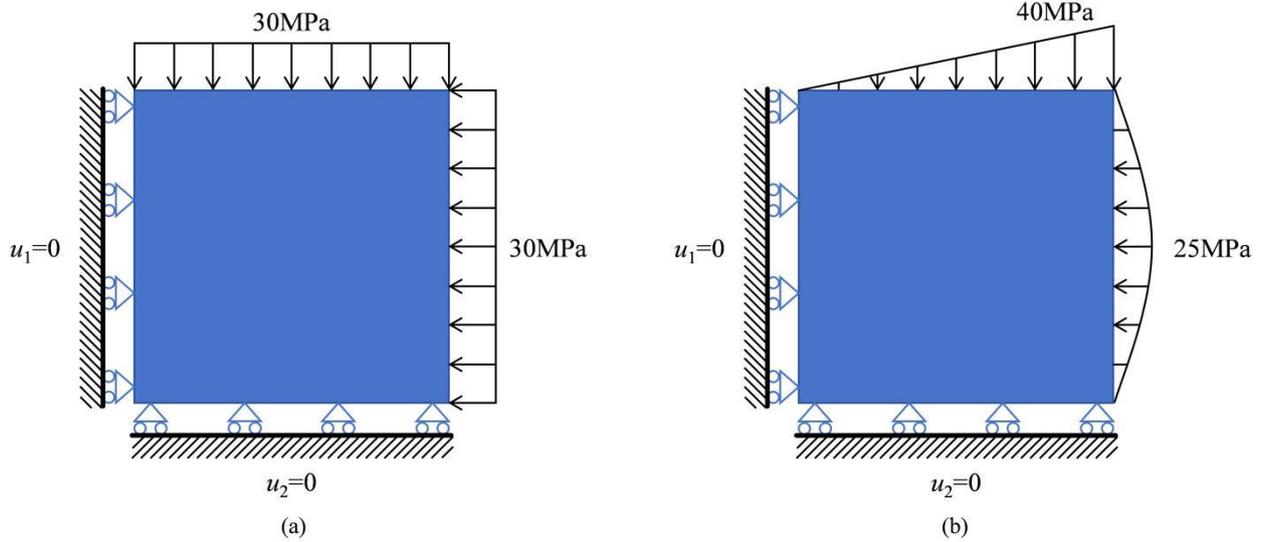

**Fig. 7.** Problem setup and boundary conditions.

### 3.2. Results of stress prediction based on MC U-net

The dataset consists of TRM microstructure images generated by cutting random fields and stress results calculated using finite element methods. All images have a resolution of 128×128. The dataset contains 1,000 sets of data, divided into training and testing sets at a ratio of 4:1. We used Attention U-net and MC U-net to predict the stress in the TRM microstructures. Both models were trained for 100 epochs, and the resulting predictions were compared with finite element results to obtain the corresponding errors. **Fig. 8a** and **Fig. 8b** show the microstructure images of TRM used for model testing and the corresponding phase interface information images. **Fig. 8c** shows the stress results obtained from finite element calculations. **Fig. 8d** and **Fig. 8e** show the stress results predicted using Attention U-net and MC U-net, respectively. **Fig. 8f** and **Fig. 8g** show the errors between the finite element results and the stress results predicted using Attention U-net and MC U-net. **Fig. 8h** shows the regions where the errors in the MC U-net predictions are reduced compared to those in the Attention U-net predictions (indicated in purple).

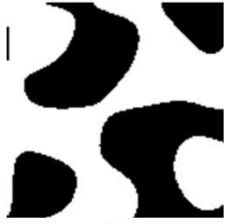
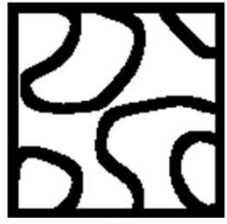

Boundary width: 6

(a)  (b)

$\sigma_{xx}$  $\sigma_{yy}$  $\sigma_{xy}$

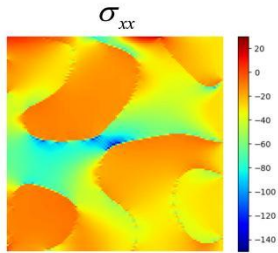
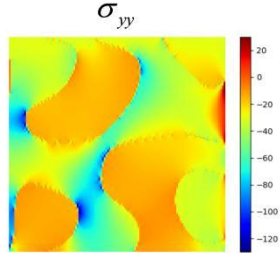
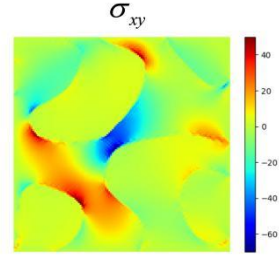

(c) FEM

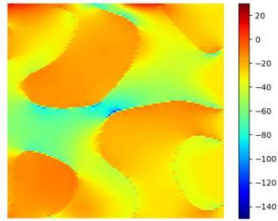
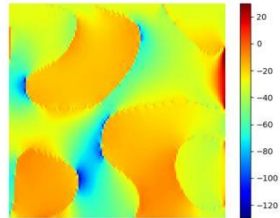
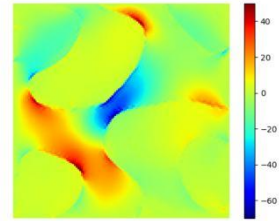

(d) Attention U-net

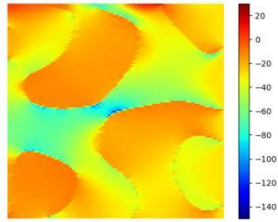
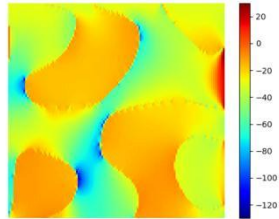
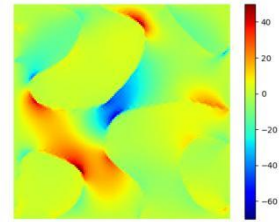

(e) MC U-net

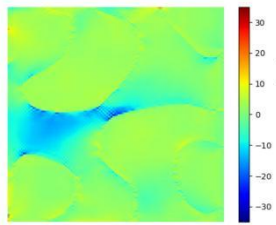 Mean error: 3.01 Max error: 31.25
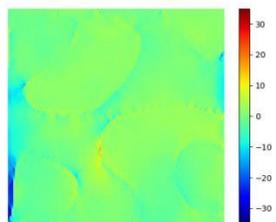 Mean error: 1.76 Max error: 31.11
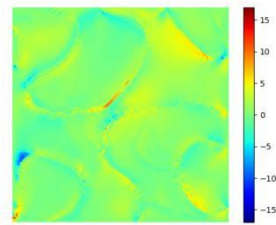 Mean error: 1.10 Max error: 16.10

(f) Error (FEM - Attention U-net)

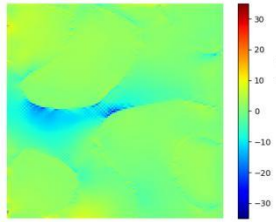 Mean error: 2.44 Max error: 27.57
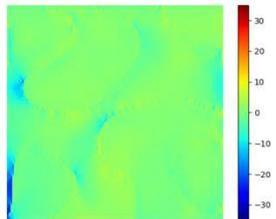 Mean error: 1.42 Max error: 27.51
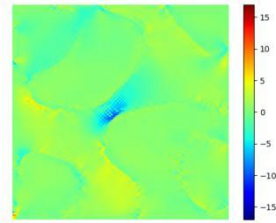 Mean error: 0.97 Max error: 11.11

(g) Error (FEM - MC U-net)

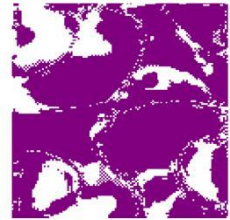 Purple pixel proportion: 72.12%
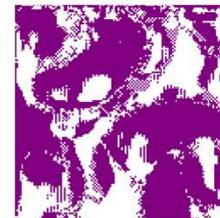 Purple pixel proportion: 61.47%
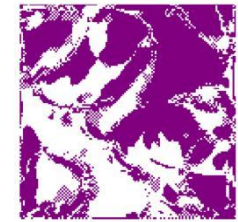 Purple pixel proportion: 56.22%

(h) Area of decreasing error (indicated in purple)

**Fig. 8.** Prediction of stress from TRM microstructure. (a) TRM microstructure. (b) Corresponding phase interface information. (c) FEM. (d) Attention U-net. (e) MC U-net. (f) Error of Attention U-net. (g) Error of MC U-net. (h) Area of decreasing error.

From the error maps of stress prediction, it can be observed that MC U-net has a more accurate prediction of stresses in the phase interface region compared to Attention U-net. This indicates that MC U-net can more precisely capture the stress concentration information of materials, which is crucial for stress prediction in material design processes. To further investigate the prediction performance of MC U-net on stress concentration regions, we considered different concatenation positions (CP) and widths of the phase interface information (WPB). We statistically analyzed the average of the mean error (AMEE) and the average of the max error (AMAE) of all stress prediction results on the test set and compared them with the results of Attention U-net, as shown in **Table 1**, where the percentage represents the ratio of the results obtained from MC U-Net to those obtained from the corresponding Attention U-Net. **Fig. 9** illustrates the different widths of the phase interface information.

Table 1. Comparison of test set prediction results under different training conditions.

| Architecture (CP) | Train hours | WPB | AMEE | | | AMAE | | |
|---|---|---|---|---|---|---|---|---|
| | | | $\sigma_{xx}$ | $\sigma_{yy}$ | $\sigma_{xy}$ | $\sigma_{xx}$ | $\sigma_{yy}$ | $\sigma_{xy}$ |
| Attention U-net | 0.56 | - | 2.17 | 2.03 | 1.59 | 26.44 | 27.14 | 14.68 |
| MC U-net (①~⑩) | 22.35 | 2 | 2.12 (98.1%) | 2.30 (113.2%) | 1.65 (103.7%) | 26.97 (102.0%) | 28.49 (104.9%) | 14.20 (96.7%) |
| | | 8 | 2.53 (117.0%) | 2.04 (100.5%) | 1.91 (120.2%) | 32.28 (122.1%) | 27.25 (100.4%) | 14.79 (100.7%) |
| MC U-net (⑥~⑩) | 11.96 | 2 | 2.02 (93.2%) | 2.23 (109.6%) | 1.64 (102.9%) | 27.33 (103.4%) | 29.26 (107.8%) | 15.40 (104.8%) |
| | | 8 | 2.17 (100.1%) | 2.01 (99.0%) | 1.93 (121.4%) | 29.16 (110.3%) | 28.08 (103.4%) | 18.03 (122.8%) |
| MC U-net (①~⑤) | 0.95 | 2 | 1.83 (84.6%) | 1.78 (87.7%) | 1.50 (94.7%) | 25.88 (97.9) | 24.64 (90.8%) | 13.04 (88.8%) |
| | | 4 | 1.90 (87.7%) | 2.06 (101.4%) | 1.43 (90.1%) | 25.97 (98.2%) | 26.70 (98.4%) | 12.62 (85.9%) |
| | | 6 | 1.87 (86.4%) | 1.72 (84.5%) | 1.45 (91.5%) | 25.39 (96.1%) | 24.51 (90.3%) | 12.59 (85.7%) |
| | | 8 | 1.95 (90.1%) | 1.84 (90.4%) | 1.58 (99.6%) | 25.62 (96.9%) | 25.07 (92.4%) | 14.27 (97.2%) |
| MC U-net (⑤) | 0.69 | 2 | 1.84 (85.2%) | 1.87 (91.9%) | 1.52 (95.8%) | 25.28 (95.63) | 24.78 (91.3%) | 13.53 (92.1%) |

| | | 4 | 1.92 (88.6%) | 1.89 (93.0%) | 1.56 (98.4%) | 25.31 (95.72%) | 25.37 (93.47%) | 13.48 (91.8%) |
|---|---|---|---|---|---|---|---|---|
| | | 6 | 2.02 (93.5%) | 2.24 (110.1%) | 1.50 (94.4%) | 26.02 (98.43) | 25.72 (94.7%) | 13.19 (89.8%) |
| | | 8 | 1.92 (88.6%) | 1.86 (91.4%) | 1.47 (92.6%) | 25.40 (96.1%) | 27.33 (100.7%) | 12.96 (88.2%) |

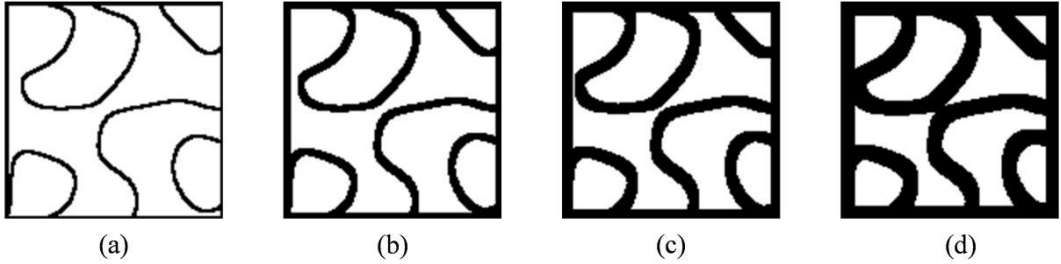

**Fig. 9.** Different widths of phase interface information in TRM microstructures. (a) 2. (b) 4. (c) 6. (d) 8.

It can be seen from **Table 1** that excessive embedding of phase interface information (excessive concatenation) significantly increases the training time of the MC U-net, while also leading to an increase in both AMEE and AMAE of the predicted stress. This indicates that excessive embedding of phase interface information disrupts the latent space of the original TRM microstructure images, which is detrimental to stress prediction. Compared to the upsampling process, embedding phase interface information during the downsampling process not only improves training efficiency but also enables more accurate stress prediction of the TRM microstructure. In contrast, embedding phase interface information at only one location during the upsampling process slightly reduces the training time, but generally results in larger prediction errors for stress, indicating that insufficient phase interface information was provided. Moreover, the width of the phase interface information should also be controlled. Excessively wide phase interface information (width=8) is detrimental to the training of the MC U-net. When the width of the phase interface information is between 2 and 6, there is no significant change in stress prediction performance. As shown in **Table 1,** the overall prediction performance of stress is optimal when embedding phase interface information during the downsampling process with an interface width of 6. Therefore, we use the stress prediction values under this condition for subsequent stress image super-resolution.

**3.3. Results of stress image super-resolution based on MPINN**

The observation data for MPINN are derived from the predictions of the MC U-net, consisting of a total of 16,384 observation points. These observation points are exclusively used to train the MPINN. Compared to stress calculations by MPINN without observation points, the number of selected points is reduced by an order of magnitude. Following the work of Ren et al. [22], the MPINN architecture includes two independent FFNNs with 12 hidden layers and 64 neurons per layer to approximate the displacement fields $u_x$ and $u_y$. Additionally, three independent FFNNs with 16 hidden layers and 64 neurons per layer are used to approximate the stress fields ($\sigma_{xx}$, $\sigma_{yy}$, $\sigma_{xy}$). The Swish activation function is utilized. The number of training epochs for MPINN is set to 1000.

After completing the training, we first performed self-prediction on all observation points, generating stress images with a resolution of 128×128 pixels. These images were then compared with the prediction results of the MC U-net and the calculation results of FEM, as shown in **Fig. 10**.

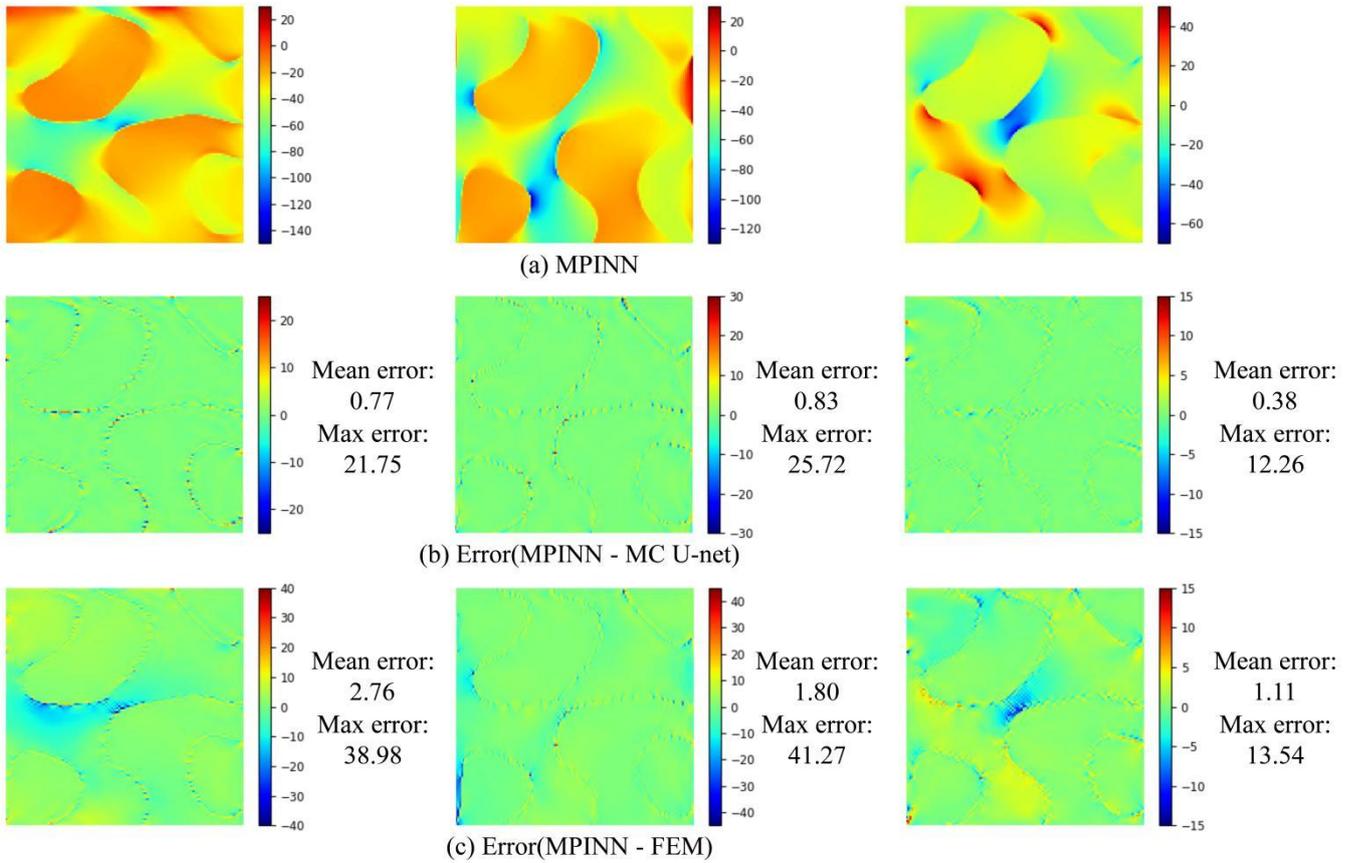

**Fig. 10.** The self-prediction results of MPINN.

Since our proposed method can upscale the resolution of existing stress images to any multiple, we investigated the relationship between the resolution enhancement effect and the magnification factor. Taking the stress image of $\sigma_{xy}$ as an example, **Fig. 11** shows the results of resolution magnification by 2 times, 4 times, 8 times, and 16 times, resulting in stress-enhanced images with resolutions of 256×256, 512×512, 1024×1024, and 2048×2048, respectively.

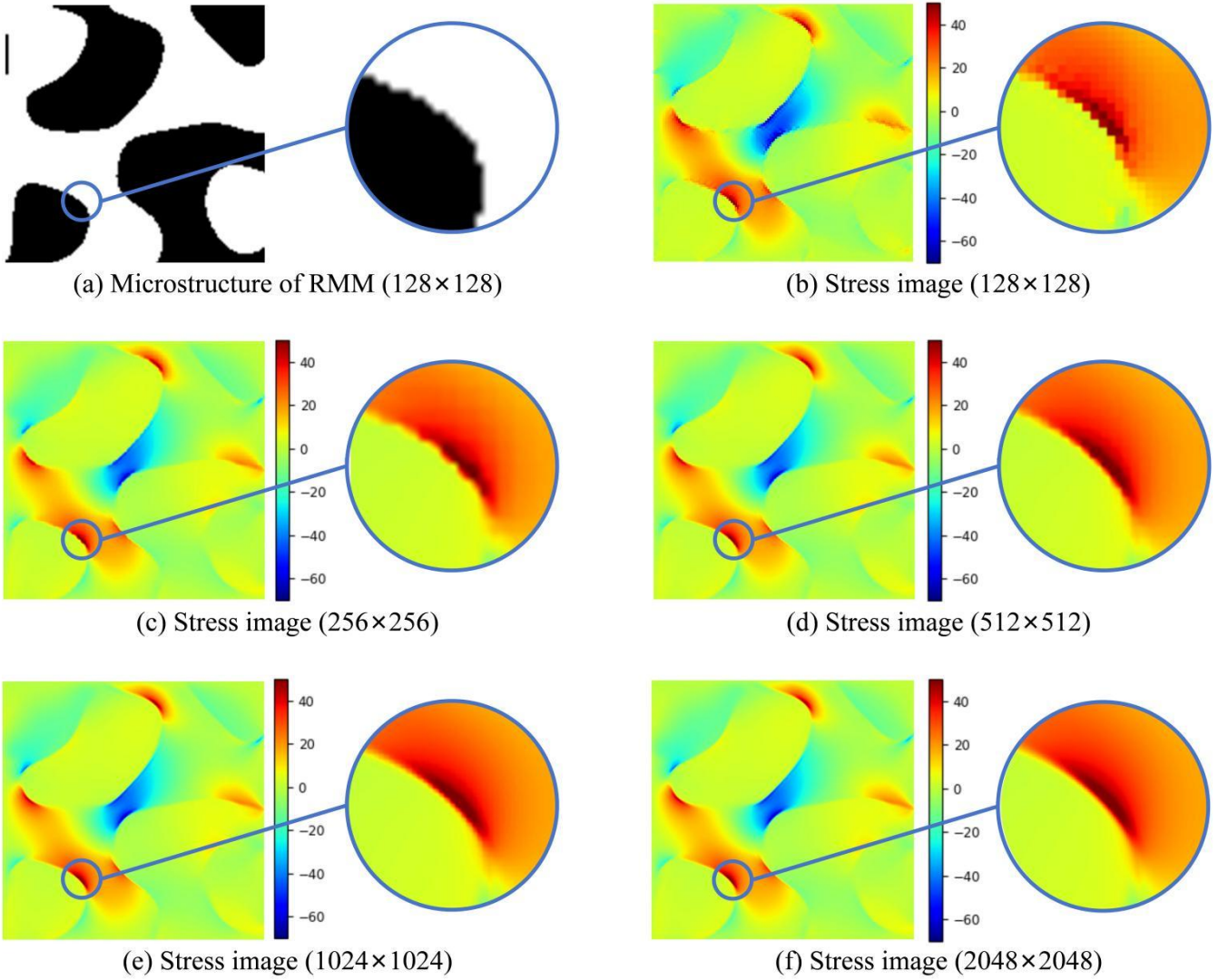

**Fig. 11.** Results of SRMPINN ($\sigma_{xy}$). (a) TRM microstructure. (b) 128×128. (c) 256×256. (d) 512×512. (e) 1024×1024. (f) 2048×2048

Considering that different weights for the the observation points part and physical information part in the loss function might lead to different prediction results (i.e., $\lambda_{PI}$ and $\lambda_{OP}$ in **Eq. (21)**). **Table 2** compares the stress results of SRMPINN with those of FEM under various weight settings.

**Table 2.** Comparison of stress results between SR-MPINN and FEM under different weighting conditions.

| Weight ($\lambda_{OP} : \lambda_{PI}$) | Resolution ratio | Mean error | | | Max error | | |
|---|---|---|---|---|---|---|---|
| | | $\sigma_{xx}$ | $\sigma_{yy}$ | $\sigma_{xy}$ | $\sigma_{xx}$ | $\sigma_{yy}$ | $\sigma_{xy}$ |
| 1 : 1 | 128×128 | 3.13 | 1.91 | 1.14 | 57.01 | 45.55 | 15.15 |
| | 256×256 | 3.17 | 1.90 | 1.16 | 85.17 | 80.77 | 32.49 |
| | 512×512 | 3.20 | 1.87 | 1.17 | 115.06 | 127.48 | 49.88 |
| | 1024×1024 | 3.23 | 1.91 | 1.18 | 152.22 | 151.67 | 64.82 |
| | 2048×2048 | 3.24 | 1.88 | 1.19 | 155.72 | 165.76 | 77.50 |
| 1:2 | 128×128 | 2.76 | 1.80 | 1.11 | 38.98 | 41.27 | 13.54 |
| | 256×256 | 2.83 | 1.81 | 1.14 | 81.62 | 73.16 | 33.52 |
| | 512×512 | 2.86 | 1.78 | 1.16 | 112.64 | 120.04 | 47.83 |
| | 1024×1024 | 2.90 | 1.83 | 1.18 | 155.61 | 149.02 | 61.95 |
| | 2048×2048 | 2.90 | 1.76 | 1.17 | 156.70 | 167.60 | 75.86 |
| 1:5 | 128×128 | 2.74 | 1.81 | 1.11 | 37.74 | 40.72 | 15.71 |
| | 256×256 | 2.82 | 1.78 | 1.14 | 83.18 | 71.77 | 34.38 |
| | 512×512 | 2.85 | 1.75 | 1.15 | 119.12 | 119.96 | 48.43 |
| | 1024×1024 | 2.90 | 1.79 | 1.17 | 154.30 | 150.06 | 62.95 |
| | 2048×2048 | 2.90 | 1.79 | 1.18 | 159.02 | 165.16 | 74.27 |
| 1:10 | 128×128 | 2.73 | 1.77 | 1.08 | 39.27 | 41.31 | 13.58 |
| | 256×256 | 2.80 | 1.77 | 1.12 | 84.28 | 75.63 | 33.63 |
| | 512×512 | 2.83 | 1.74 | 1.37 | 118.63 | 126.68 | 50.51 |
| | 1024×1024 | 2.88 | 1.79 | 1.16 | 149.25 | 155.66 | 63.05 |
| | 2048×2048 | 2.87 | 1.76 | 1.16 | 153.45 | 170.21 | 75.90 |

  We found that when the proportion of physical information in the loss function is too low, the error compared to the FEM results is larger. This is because MPINN is trained using the stress prediction results from MC U-net, and the constraints of the observation points can cause some error accumulation. However, if the proportion of the physical information part in the loss function is increased, MPINN tends to reduce the loss caused by the physical information part, which helps to correct the accumulated errors to some extent. Nonetheless, as the proportion of the observation points part in the loss function gradually decreases, some maximum errors increase. This is because when the loss function is fully controlled by physical information, a large number of training points are required [22]. Therefore, we recommend a weight ratio of 1:5 ($\lambda_{OP} : \lambda_{PI} = 1:5$). Additionally, we observed that the mean error and maximum error of SRMPINN stress results gradually increase with higher resolution. However, when the resolution increases to 1024×1024, the mean error and maximum error almost no longer change. This indicates that the error in SRMPINN stress results converges.

  To demonstrate the superiority of our proposed ISR method, **Fig. 12** compares the enhancement effects with SRGAN at 4x and 16x magnification factors. The weight ratio of the loss function in SR-MPINN is set to 1:5 ($\lambda_{OP} : \lambda_{PI} = 1:5$). The regions indicated in purple show that the errors in the SRMPINN predictions are reduced compared to those in the SRGAN predictions.

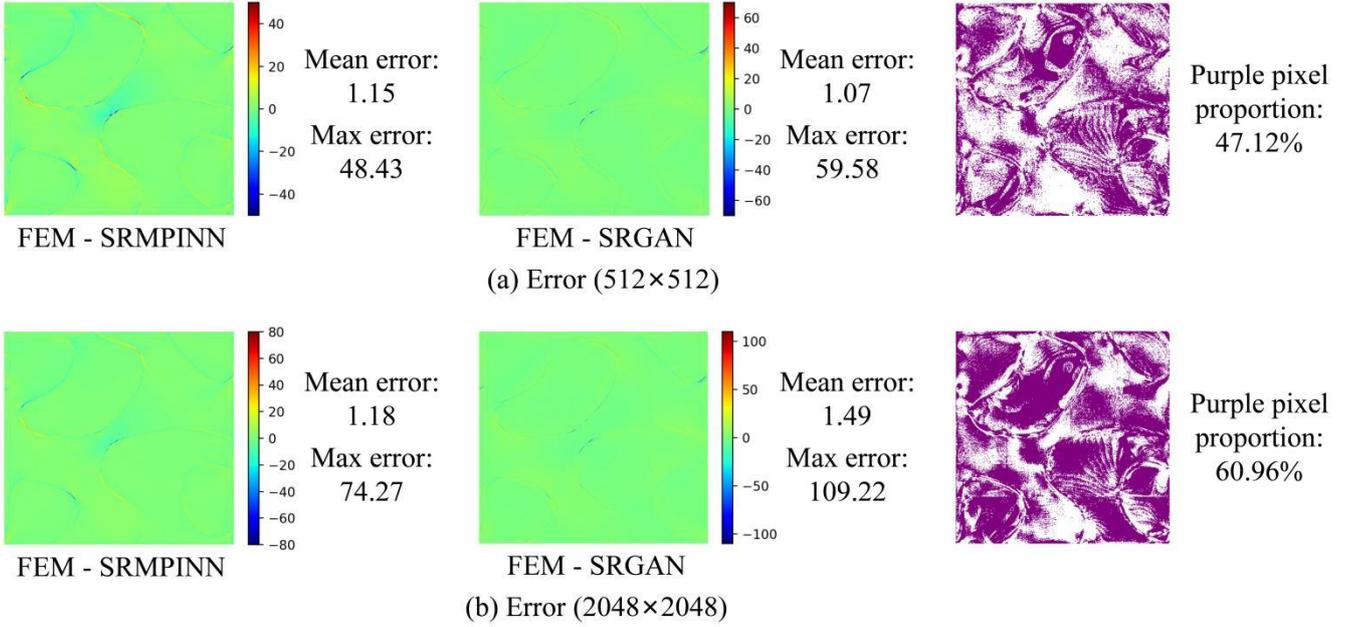

**Fig. 12.** Comparison of super-resolution results between SRMPINN and SRGAN.

We observed that when the image magnification factor is small, SRMPINN and SRGAN perform similarly, with SRGAN demonstrating advantages in certain evaluation metrics. However, when the image magnification factor is large, SRMPINN exhibits significant advantages in all evaluation metrics. This can be explained using the results presented in Table 2: as the image magnification factor increases, the error of SRMPINN changes slowly and gradually converges. However, SRGAN lacks this characteristic. From Fig. 12, we can observe that as the resolution changes from 512×512 to 2048×2048, the mean error and max error of SRGAN increase by nearly 1.5 times and 2 times, respectively. It is worth mentioning that SRMPINN is not a fully data-supervised learning method, which is a significant advantage compared to data-driven ISR methods such as SRGAN.

## 4. Transfer learning

To demonstrate the generalization ability of the proposed stress analysis framework, we developed a transfer learning model in this section to validate its performance on new datasets. In the examples of this section, we first changed the volume fractions of different phases in the TRMs. Then, the biaxial uniform compression was modified to biaxial non-uniform compression, and the applied uniform pressure was changed to non-uniform pressure, as shown in Fig. 7b. We used the model weights from the original task as the initial weights for the new task, leveraging prior knowledge from Section 3.

Fig. 13 shows the loss functions of the two network architectures, MC U-net (using $\sigma_{xx}$ as an example) and SR-MPINN, during transfer learning. It was found that using the pre-trained model from Section 3 as the initial training model for the new problem accelerates the convergence speed of the neural network and results in lower loss function values.

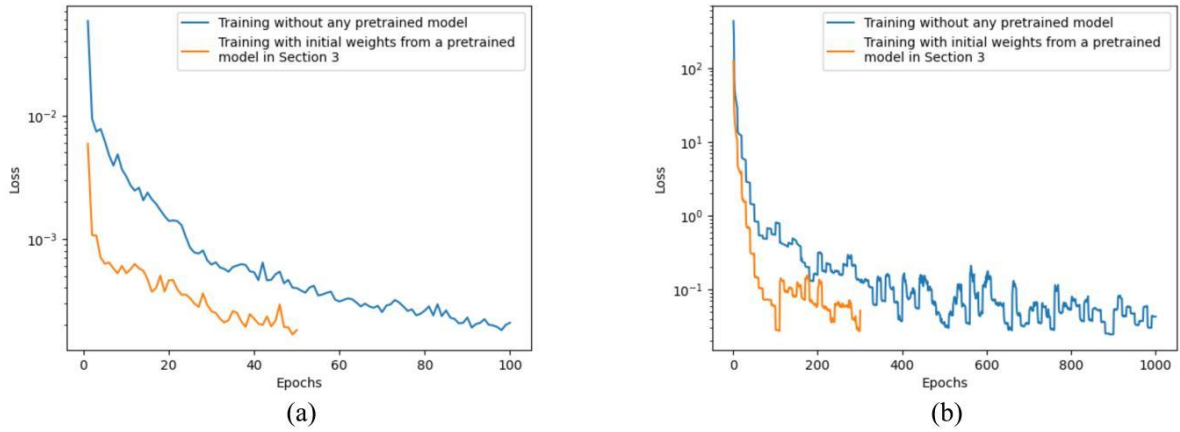

**Fig. 13.** Effect of transfer learning on the total loss. (a) MC U-net ($\sigma_{xx}$). (b) MPINN

Fig. 14 shows the stress prediction results obtained by using the pre-trained MC U-net model from Section 3 as the initial model for training. Fig. 15 shows the stress image super-resolution results obtained by using the pre-trained SR-MPINN model from Section 3 as the initial model for training (using $\sigma_{xy}$ as an example). The stress-enhanced images with resolutions of 128×128, 256×256, 512×512, 1024×1024, and 2048×2048 are presented.

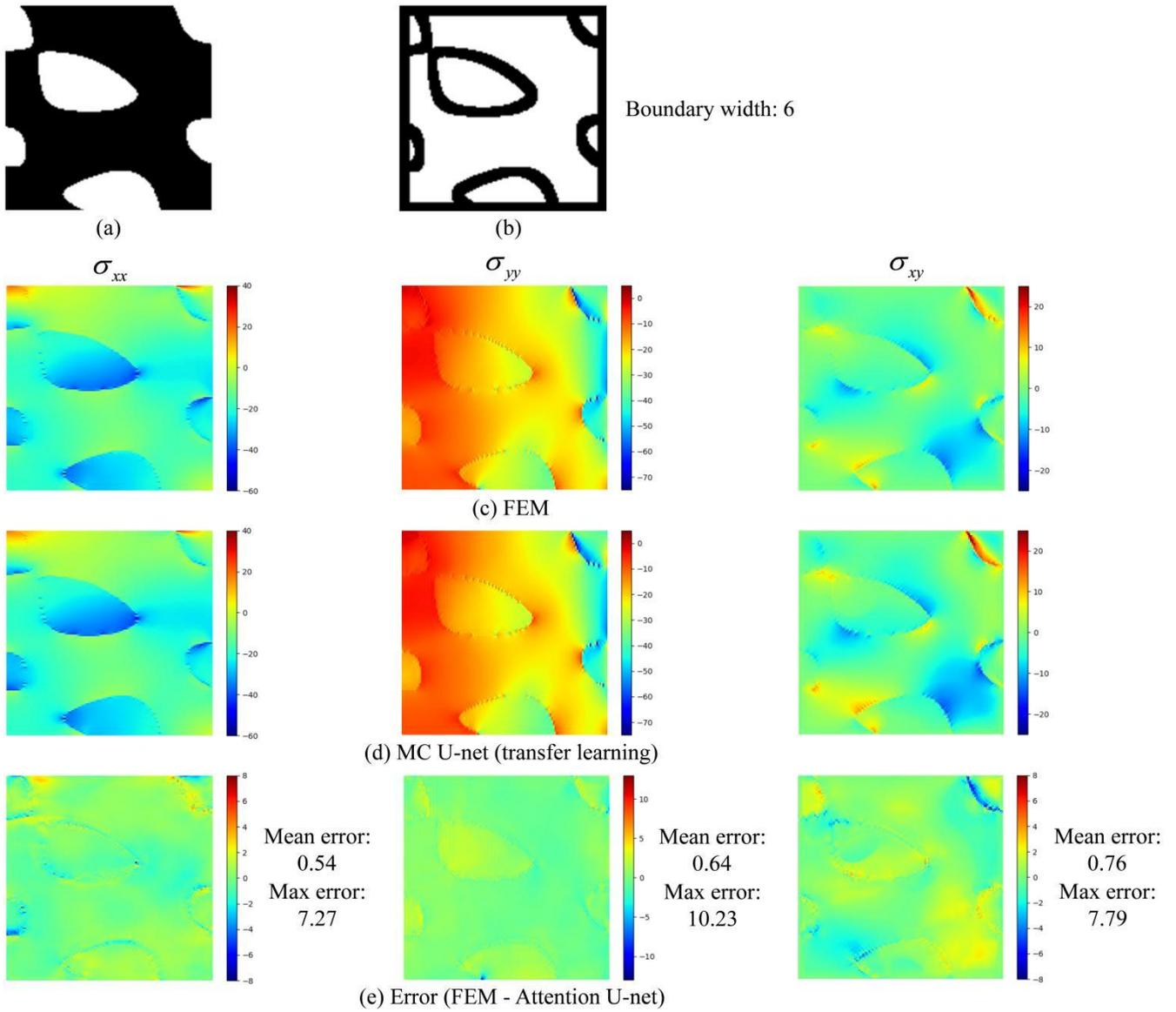

**Fig. 14.** Prediction of stress from transfer learning. (a) TRM microstructure. (b) Corresponding phase interface information. (c) FEM. (d) MC U-net. (e) Error of MC U-net.

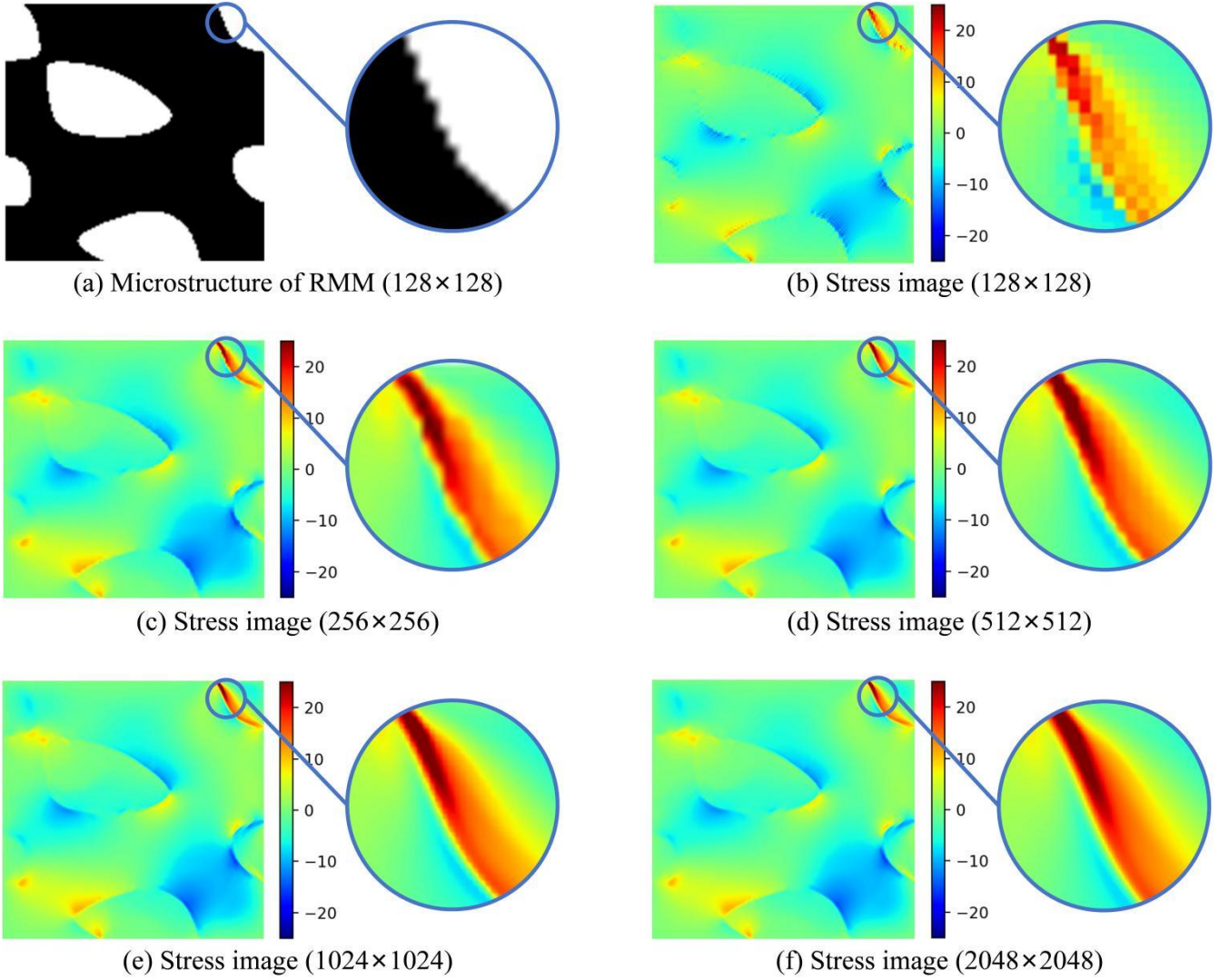

**Fig. 15.** Results ($\sigma_{xy}$) of SRMPINN from transfer learning. (a) TRM microstructure. (b) 128×128. (c) 256×256. (d) 512×512. (e) 1024×1024. (f) 2048×2048

The above results demonstrate the generalization ability of our proposed stress prediction framework, which can predict the stress of different TRMs under various loading conditions. This provides a new tool for quickly obtaining the stress distribution of materials and focusing on stress concentration areas in the material inverse design.

## 5. Conclusions

In this study, we developed a stress analysis framework for two-phase random materials (TRMs). This framework can be used for stress analysis of materials with almost any form of microstructure. The conclusions of the specific content are as follows:

1. We generated random fields using random harmonic functions and performed horizontal slicing on

the random fields to obtain TRM microstructure images.

2. We proposed the Multiple Compositions U-net (MC U-net) architecture, which fully considers the phase interface information of TRM microstructures, including the concatenation positions and the width of the phase interface information. We recommend concatenating only during the downsampling process and selecting phase interface widths between 2 and 6, effectively reducing the stress prediction errors at the phase boundaries. Compared to the Attention U-net, various evaluation metrics are reduced by approximately 5% to 15%.

3. We proposed a stress ISR method based on MPINN (SRMPINN). This method only requires the stress prediction results from MC U-net for its training set. Due to the presence of observation points, SRPINN requires only about 16k points. This is an order of magnitude fewer points compared to previous studies using MPINN for stress prediction. By using the stress values of the corresponding observation points and the constraints of physical information, the resolution of the stress images can be upscaled to any multiple. We investigated the relationship between the resolution magnification and the enhancement effect on the stress images, finding that the prediction errors converge. Additionally, we considered the impact of the weight ratio between the physical information part and the observation points part in the loss function on the stress image enhancement effect and provided a recommended weight ratio ($\lambda_{\text{OP}} : \lambda_{\text{PI}} = 1:5$). Subsequently, we compared the effects of our proposed method with those of SRGAN, showing that as the image magnification factor increases, SRMPINN exhibits significant advantages in all evaluation metrics. Our proposed SRMPINN method breaks the limitation of grid accuracy on stress images, allowing us to train the MC U-net model using material microstructure images with lower resolution and obtain stress images with any resolution.

4. We conducted transfer learning on TRMs with different phase volume fractions and different loading conditions, demonstrating the generalization capability of our proposed stress analysis framework.

In conclusion, this framework integrates the construction of TRM microstructures, stress prediction, stress image super-resolution, and transfer learning into a cohesive approach, as shown in **Fig. 1**. Compared to previous stress prediction methods, it not only offers more precise and efficient predictions for stress concentration areas when dealing with complex problems but also enables a multiscale analysis of the stress concentration regions at phase boundaries. This endows deep learning-based stress prediction methods with greater potential for engineering applications. However, the framework is still not perfect for solving practical three-dimensional problems, dynamic problems, and material fracture problems. Future research can focus on these challenges to further enhance the adaptability of this stress analysis framework in practical engineering problems.